\documentclass[11pt,nofootinbib,floats,article,aps,prd,floatfix,titlepage,superscriptaddress]{revtex4} 
\usepackage{epsfig}
\usepackage{graphics}
\usepackage{bm}
\usepackage{amsmath}

\def\({\left(}
\def\){\right)}
\def\[{\left[}
\def\]{\right]}

\def\nn{\nonumber}
\def\ee{\end{equation}}
\def\eea{\end{eqnarray}}
\def\nn{\nonumber \\}
\def\p{\partial}
\def\be{\begin{equation}}
\def\bea{\begin{eqnarray}}
\def\h{\frac12}

\begin{document}

\title{State of matter at high density and entropy bounds}\footnote{Essay written for the Gravity Research Foundation 2015 Awards for Essays on Gravitation.}

\author{Ali Masoumi}
\email{ali@cosmos.phy.tufts.edu}
\affiliation{Tufts Institute of Cosmology,
255 Robinson Hall, 212 College Ave,
Medford, MA, 02155
}

\begin{abstract}

Entropy of all systems that we understand well is proportional to their volumes except for black holes given by their horizon area. This makes the microstates of any quantum theory of gravity drastically different from the ordinary matter. Because of the assumption that black holes are the maximum entropy states there have been many conjectures that put the area, defined one way or another, as a bound on the entropy in a given region of spacetime. Here we construct a simple model with entropy proportional to volume which exceeds the entropy of a single black hole. We show that a homogeneous cosmology filled with this gas exceeds one of the tightest entropy bounds, the covariant entropy bound and discuss the implications. 

\end{abstract}

\maketitle

Bekenstein's discovery \cite{bek}  of black hole entropy and Hawking's calculation \cite{Hawking:1974sw}  of the temperature are among the most important clues we have on a quantum theory of gravity. These show a significant difference between the microstates  of quantum gravity and ordinary matter. Quantum field theory, the way we understand it, implies that the density of states in momentum space is proportional to the volume and so is the entropy. 

The assumption, or the bias, that black holes are the highest entropy states brought about many conjectures on the maximum entropy confined  in a region. In the absence of gravity, one can put arbitrary entropy in a region by inserting enough energy in it. Gravity changes this picture. Energy distorts the geometry and volume in such a way that the entropy is no longer arbitrarily large. Bekenstein \cite{bek2} conjectured that in a system where gravity is not the dominant interaction, the entropy in a region is bounded by the entropy  of a black hole of the same size. This conjecture fails  in well known examples. Consider  a cosmology with flat spatial slices and ordinary matter, like radiation. The entropy  is proportional to the volume which grows much faster than the area and exceeds it for large enough regions. Later, several more refined versions of this conjecture were stated \cite{fs, bousso} that  survive in more general circumstances.

 All of these conjectures assume some constraint on the   energy-momentum tensor like the weak  or dominant energy conditions. It is not possible to put any bound on entropy if one allows  an arbitrary energy-momentum tensor. For example, consider  a combination of positive and negative energy states with a zero net energy-momentum. These states would not have any feedback on geometry and hence, can violate any entropy bound. Therefore, in a sense, the entropy bounds are limitations on the form of energy-momentum tensor rather than restricting the entropy. Putting  enough restrictions on the energy-momentum tensor, one can  prove some of the bounds in weak gravity limit \cite{fmw, bfm2} . However, we should be cautious about putting too much constraint on the states of a theory (quantum gravity) that we do not understand well.
 
  Here we motivate a heuristic  model which satisfies the dominant energy condition and breaks one of the tightest entropy bounds, the covariant (Bousso)  bound \cite{bousso}. The details and the subtleties of this model is discussed  in \cite{alimathur,Masoumi:2014nfa}.
 
\begin{figure}[htbp] 
    \centering
    \includegraphics[width=3.4in]{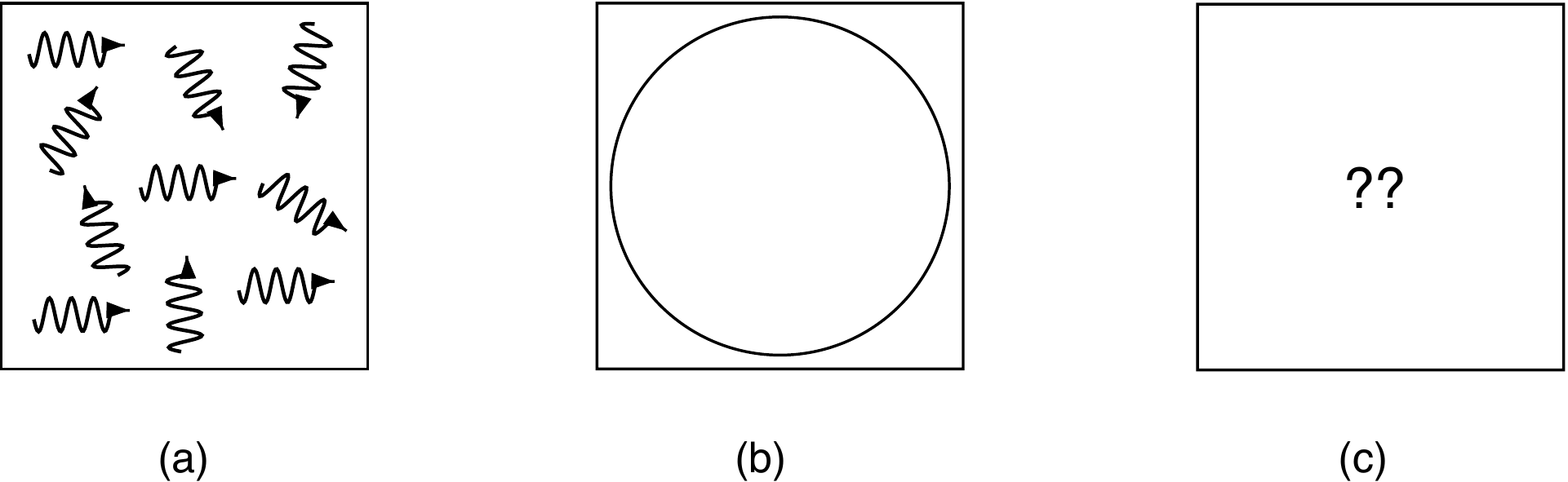} 
    \caption{For small densities we expect the maximum entropy state to be a gas of relativistic particles (left), increasing the energy we expect the matter to collapse into a black hole (middle). Does a phase with higher energy and entropy exist(right)? }
    \label{fig:Entorpy}
\end{figure}

\begin{figure}[htbp] 
    \centering
    \includegraphics[width=2.2in]{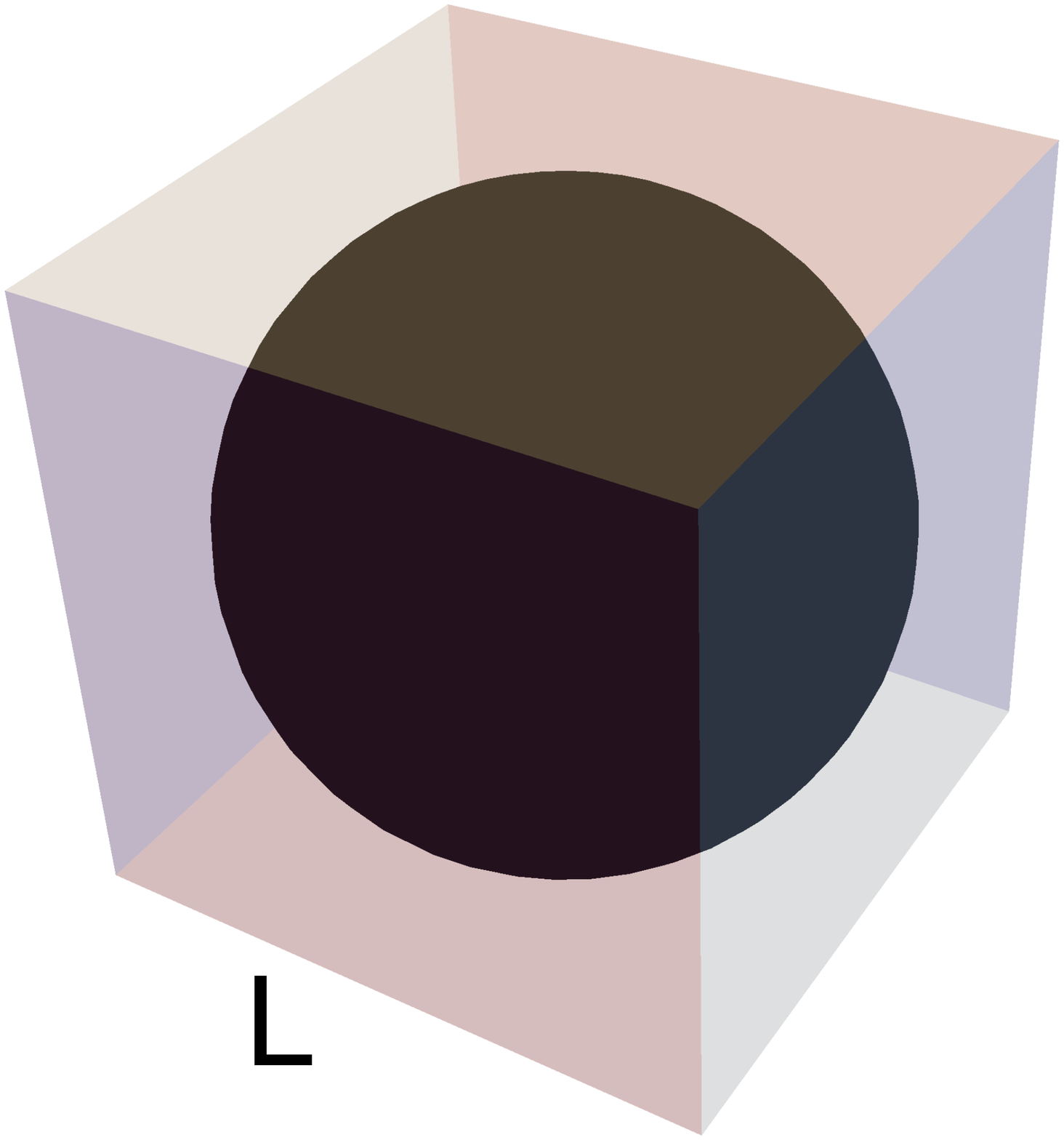} 
    \includegraphics[width=2.2in]{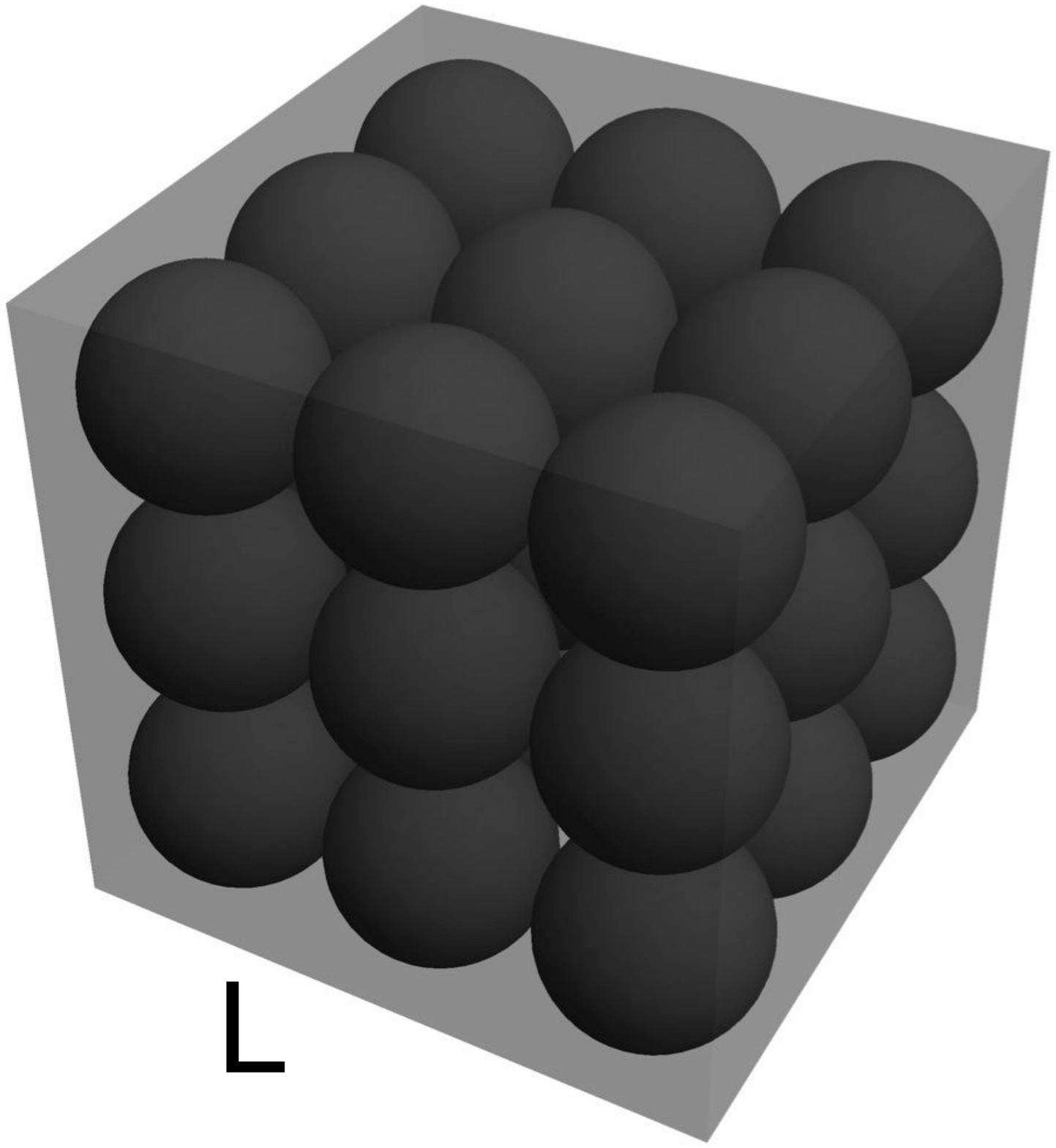} 
    \caption{Left, a box of size $L$ contains a single black hole. Right, the same box, filled with 27 smaller black holes of size $L/3$. The combined horizon area of smaller black holes is 3 times larger than the left panel.}
    \label{fig:SingleVsLattice}
\end{figure}

Let's investigate the maximum entropy one can put in a cube  of size $L$. As depicted in Fig.\ref{fig:Entorpy}, at small energy densities we expect a gas of relativistic particles to have the highest entropy. Increasing the energy in the box  makes it collapse into a black hole which  indeed has a much higher  entropy. Can we put more entropy and energy in this box?  The answer is yes and a simple configuration is motivated in Fig.\ref{fig:SingleVsLattice}. Each of the 27 smaller holes on the right panel have  1/9 of the area and entropy of the  hole on the left.
Hence, the net entropy of the right panel is  3 times of the single hole configuration. 

Using $N^3$ holes of size $L/N$, we get  $N$ times more entropy than a single black hole. If this box is put in empty space, it is entropically favorable for the smaller ones to merge into a single black hole. The result is a black hole of size $N^2 L$ and $N^5$ times the  entropy of the original lattice. However, if the whole universe is filled with this black hole lattice, there is no room for them to merge. It is entropically favorable to have this lattice until the universe expands enough and they can merge then.

We  may worry that this system is completely  unstable to gravitational collapse. However, to surpass the single black hole entropy we do not necessarily  create  strong gravitational fields.  One can build a gas of  small black holes where each  is well outside the  Schwarzschild  radius of the others. Suppose the separation between the black holes is $n$ times larger than their Schwarzschild radius. To exceed the entropy of the single black hole configuration, we need small black holes of size $R \le L/n^3$. Consider a  solar system size box in which the black holes are separated by 10000 times of their horizon radius. We will need small black holes of size 5 meters to surpass the single black hole entropy. The big separation of the black holes prevents an immediate coalescence.  
This  example shows that even using the simple ingredients of classical gravity we can get entropies  larger than the one for a single  black hole. 

 In more realistic scenarios obtained form string theory we  replace this lattice of black holes with a set of intersecting branes. These branes  interact with each other and  have pressure and thermodynamics. Let's consider a gas of these black holes with typical sizes $R_{\rm s}$ and  separation of the same order $R_{\rm s}$.  Each  black hole has   entropy $S_{\rm bh} \sim R_{\rm s}^2/G$ and energy $E_{\rm bh}\sim R_{\rm s}/G$ and there are $N_{\rm bh}\sim V/R_{\rm s}^3$ of them in a volume $V$. The total energy and entropy is given by 
\begin{eqnarray}
S\sim N_{\rm bh} S_{\rm bh}\sim {V\over R_{\rm s} G}~, \qquad E\sim N_{\rm bh} E_{\rm bh}\sim {V\over R_{\rm s}^2 G}~.
\label{entropylattice}
\end{eqnarray}
Eliminating $R_{\rm s}$ between these two equations we get 
\be
S\sim {1\over R_s} {V\over G}\sim \left ( {V\over EG}\right )^{-\h}{V\over G} \sim \sqrt{EV\over G}
\ee
Introducing a constant $K$ of order unity and rewriting in terms of energy and entropy densities $\rho$ and $s$ we get 
\be \label{EntropyFormul}
s=K\sqrt{\rho\over G}~.
\ee
This gives an extensive equation for entropy. Using the first law of thermodynamics $dE = T dS - p d V$ we get 
\begin{eqnarray}
  T&=&\left ( {\p S\over \p E}\right ) _V^{-1}={2\over K} \sqrt{EG\over V}~, \nn 
  p&=&T\left ( {\p S\over \p V }\right ) _E = {E\over V}=\rho~.
  \label{Thermo}
\end{eqnarray}
This corresponds to the equation of state $p=\rho$ which satisfies the dominant energy condition. Although we derived Eq.\eqref{EntropyFormul} using a heuristic model, this equation has been derived from many different approaches including: 
\begin{enumerate}
	\item Veneziano \cite{veneziano2} obtained it  assuming that the entropy is given by the area of the apparent horizon with radius  $H^{-1} \sim \(G \rho\)^{-\frac12}$.
	\item  Banks and Fishcler \cite{bf} derived $p=\rho$ using coalescing horizon size black holes.  
	\item Horowitzand and Polchinski \cite{hp} derived it for a gas of string states near the point where the strings are large enough to collapse into a black hole. 
 	\item Sasakura \cite{sas1} conjectured it  by demanding  a `spacetime uncertainty relation'.   
	\item Verlinde  \cite{verlinde}  argued that a similar relation  correspond to the Cardy formula for the density of states. 
	\item Brustein and Veneziano  \cite{brusv} obtained it  by using the notion  of a causal connection. 
	 \item Masoumi and Mathur  \cite{alimathur} showed that Eq.\eqref{EntropyFormul} is the only entropy  formula which is manifestly invariant under S and T-dualities and approaches the black hole entropy at small densities.
 
\end{enumerate}
Now we show that this  high-entropy 'matter' can violate one of the most beautiful entropy  bounds, the covariant (Bousso) bound.  This bound  is based on the entropy in the light sheet and has a very different philosophy and formulation. The question of interest in the earlier bounds was: How much is the maximum entropy one can put in a given region? The answers were typically the area of that region in Planck units. The covariant entropy bound changes the question: Given a two-dimensional surface $\mathcal S$ of area $A$, what is the region which has a maximum entropy given by $A$? The conjecture  was expressed in terms of non-expanding null geodesic congruences. Take a bundle of null rays perpendicular to $\mathcal S$ and follow them until they reach a caustic or start diverging. The region covered by this  bundle is called the light-sheet. The conjecture expresses that the entropy of this light-sheet is bounded by $A$.  We show that  a homogeneous cosmology with entropy density given in Eq.\eqref{EntropyFormul} exceeds this bound. Assume   a Bianchi type I metric  
\begin{equation}
	ds^2= -dt^2 + \sum_{i=1}^3 a_i(t)^2 dx_i^2~.
\end{equation}
This  is a homogenous anisotropic  universe. The Friedmann equations for this universe filled with the matter discussed in Eq.\eqref{EntropyFormul} have a power law solution in the form
\begin{equation}
	a_i(t) =a_{0i} t^{C_i}~\text{for} \quad  0<C_i<1 ~.,
\end{equation}
where $\sum_{i=1}^3 C_i=1$. Let's choose a surface $\mathcal S $ in the $(x_2-x_3)$ plane. If  $S_{\rm bound}$ is the maximum entropy allowed by the covariant  bound, we can show 
\begin{equation}\label{Req}
 \text{ratio}={S_{\rm LightSheet}\over S_{\rm bound}}={K\left(1-\sum_i C_i^2\right)^\h\over \sqrt{\pi}(1-C_1)}~.
\end{equation}
This ratio can get bigger than one  for {\it any} value of $K$ if $C_1$ is close enough to one.  For example if $K=1$  we need $C_1=0.85$ which is a mild anisotropy. The ratio remains the same even if we do not follow the light sheet all the way to the initial singularity.  This is in contrast with  the conditions for ordinary matter like radiation. To break the bound for radiation we must follow the light sheet to Planckian  densities where we have no good understanding.

Here we constructed a model which violates the tightest of entropy bounds. There is no simple reason to assume this model is not allowed on first principles. If a shear viscosity prevents large anisotropy we may be able to find a deeper connection between the viscosity and entropy bounds. If this violation stands as it is, it has important consequences. The simplest would be tightening the assumptions of the area-based entropy bounds. This would render these conjectures less useful. As mentioned above, the entropy bounds are really constraints on the type of matter allowed by the theory. The bounds would not be very impressive under  too much limitation. A deeper consequence would be returning to the more natural bounds based on the volume instead of area. This will bring the number counting for the states allowed by gravity to the same footing as other quantum theories where  we have a much better understanding and control. 
\section*{Acknowledgement}
I am grateful to S. Mathur, K. Olum and E. Weinberg for the interesting discussions.


\end{document}